\documentclass[a4paper,12pt]{article}

\usepackage{latexsym}

\newcommand{\nc}{\newcommand}    

\nc{\be}[1]{\begin{equation}\mbox{$\label{#1}$}}
\nc{\bea}[1]{\begin{eqnarray} \mbox{$\label{#1}$}}
\nc{\Section}[2]{\section{#2}\label{#1}}
\nc{\Bibitem}[1]{\bibitem{#1}}
\nc{\Label}[1]{\label{#1}}

\nc{\eea}{\end{eqnarray}}
\nc{\ee}{\end{equation}}

\nc{\bdm}{\begin{displaymath}}
\nc{\edm}{\end{displaymath}}
\nc{\dpsty}{\displaystyle}
\nc{\bc}{\begin{center}}
\nc{\ec}{\end{center}}
\nc{\ba}{\begin{array}}
\nc{\ea}{\end{array}}
\nc{\bab}{\begin{abstract}}
\nc{\eab}{\end{abstract}}
\nc{\btab}{\begin{tabular}}
\nc{\etab}{\end{tabular}}
\nc{\bit}{\begin{itemize}}
\nc{\eit}{\end{itemize}}
\nc{\ben}{\begin{enumerate}}
\nc{\een}{\end{enumerate}}
\nc{\bfig}{\begin{figure}}
\nc{\efig}{\end{figure}}

\nc{\arreq}{&\!=\!&}
\nc{\arrmi}{&\!-\!&}
\nc{\arrpl}{&\!+\!&}
\nc{\arrap}{&\!\!\!\approx\!\!\!&}
\nc{\non}{\nonumber}
\nc{\align}{\!\!\!\!\!\!\!\!&&}

\def\lsim{\; \raise0.3ex\hbox{$<$\kern-0.75em
      \raise-1.1ex\hbox{$\sim$}}\; }
\def\gsim{\; \raise0.3ex\hbox{$>$\kern-0.75em
      \raise-1.1ex\hbox{$\sim$}}\; }
\def\tilde{\raise0.3ex\hbox{\kern0em
      \raise-0.2ex\hbox{$\sim$}}}

\nc{\DOT}{\hspace{-0.08in}{\bf .}\hspace{0.1in}}
\nc{\Laada}{\hbox {$\sqcap$ \kern -1em $\sqcup$}}
\nc\loota{{\scriptstyle\sqcap\kern-0.55em\hbox{$\scriptstyle\sqcup$}}}
\nc\Loota{{\sqcap\kern-0.65em\hbox{$\sqcup$}}}
\nc\laada{\Loota}
\nc{\qed}{\hskip 3em \hbox{\BOX} \vskip 2ex}

\nc{\real}{{\rm I \! R}}
\nc{\Z}{{\sf Z \!\!\! Z}}
\nc{\complex}{{\rm C\!\!\! {\sf I}\,\,}}
\def\bigid{\leavevmode\hbox{\small1\kern-3.8pt\normalsize1}}
\def\id{\leavevmode\hbox{\small1\kern-3.3pt\normalsize1}}
\nc{\slask}{\!\!\!/}
\nc{\bis}{{\prime\prime}}
\nc{\pa}{\partial}
\nc{\na}{\nabla}
\nc{\ra}{\rangle}
\nc{\la}{\langle}
\nc{\goto}{\rightarrow}
\nc{\swap}{\leftrightarrow}

\nc{\EE}[1]{ \mbox{$\cdot10^{#1}$} }
\nc{\abs}[1]{\left|#1\right|}
\nc{\at}[2]{\left.#1\right|_{#2}}
\nc{\norm}[1]{\|#1\|}
\nc{\abscut}[2]{\Abs{#1}_{\scriptscriptstyle#2}}
\nc{\vek}[1]{{\rm\bf #1}}
\nc{\integral}[2]{\int\limits_{#1}^{#2}}
\nc{\inv}[1]{\frac{1}{#1}}
\nc{\dd}[2]{{{\partial #1}\over{\partial #2}}}
\nc{\ddd}[2]{{{{\partial}^2 #1}\over{\partial {#2}^2}}}
\nc{\dddd}[3]{{{{\partial}^2 #1}\over
    {\partial #2 \partial #3}}}
\nc{\dder}[2]{{{d #1}\over{d #2}}}
\nc{\ddder}[2]{{{d^2 #1}\over{d {#2}^2}}}
\nc{\dddder}[3]{{d^2 #1}\over
    {d #2 d #3}}
\nc{\dx}[1]{d\,^{#1}x}
\nc{\dy}[1]{d\,^{#1}y}
\nc{\dz}[1]{d\,^{#1}z}
\nc{\dl}[1]{\frac{d\,^{#1}l}{(2\pi)^{#1}}}
\nc{\dk}[1]{\frac{d\,^{#1}k}{(2\pi)^{#1}}}
\nc{\dq}[1]{\frac{d\,^{#1}q}{(2\pi)^{#1}}}

\nc{\bfT}{{\bf T }}
\def\GeV{{\rm\ GeV}}

\nc{\cA}{{\cal A}}
\nc{\cB}{{\cal B}}
\nc{\cD}{{\cal D}}
\nc{\cE}{{\cal E}}
\nc{\cG}{{\cal G}}
\nc{\cH}{{\cal H}}
\nc{\cL}{{\cal L}}
\nc{\cO}{{\cal O}}
\nc{\cT}{{\cal T}}
\nc{\cN}{{\cal N}}
\nc{\rvac}[1]{|{\cal O}#1\rangle}
\nc{\lvac}[1]{\langle{\cal O}#1|}
\nc{\rvacb}[1]{|{\cal O}_\beta #1\rangle}
\nc{\lvacb}[1]{\langle{\cal O}_\beta #1 |}
\nc{\bb}{\bar{\beta}}
\nc{\bt}{\tilde{\beta}}
\nc{\ctH}{\tilde{\cal H}}
\nc{\chH}{\hat{\cal H}}
\nc{\1}{\aa}
\nc{\2}{\"{a}}
\nc{\3}{\"{o}}
\nc{\4}{\AA}
\nc{\5}{\"{A}}
\nc{\6}{\"{O}}
\nc{\al}{\alpha}
\nc{\g}{\gamma}
\nc{\Del}{\Delta}
\nc{\e}{\textrm{e}}
\nc{\eps}{\epsilon}
\nc{\lam}{\lambda}
\nc{\Om}{\Omega}
\nc{\ve}{\varepsilon}
\nc{\mn}{{\mu\nu}}
\nc{\vp}{\varphi}

\nc{\advp}[3]{{\it  Adv.\ in\ Phys.\ }{{\bf #1} {(#2)} {#3}}}
\nc{\annp}[3]{{\it  Ann.\ Phys.\ (N.Y.)\ }{{\bf #1} {(#2)} {#3}}}
\nc{\apl}[3]{{\it  Appl. Phys. Lett. }{{\bf #1} {(#2)} {#3}}}
\nc{\apj}[3]{{\it  Ap.\ J.\ }{{\bf #1} {(#2)} {#3}}}
\nc{\apjl}[3]{{\it  Ap.\ J.\ Lett.\ }{{\bf #1} {(#2)} {#3}}}
\nc{\app}[3]{{\it Astropart.\ Phys.\ }{{\bf #1} {(#2)} {#3}}}
\nc{\cmp}[3]{{\it  Comm.\ Math.\ Phys.\ }{{ \bf #1} {(#2)} {#3}}}
\nc{\cqg}[3]{{\it  Class.\ Quant.\ Grav.\ }{{\bf #1} {(#2)} {#3}}}
\nc{\epl}[3]{{\it  Europhys.\ Lett.\ }{{\bf #1} {(#2)} {#3}}}
\nc{\ijmp}[3]{{\it Int.\ J.\ Mod.\ Phys.\ }{{\bf #1} {(#2)} {#3}}}
\nc{\ijtp}[3]{{\it Int.\ J.\ Theor.\ Phys.\ }{{\bf #1} {(#2)} {#3}}}
\nc{\jmp}[3]{{\it  J.\ Math.\ Phys.\ }{{ \bf #1} {(#2)} {#3}}}
\nc{\jpa}[3]{{\it  J.\ Phys.\ A\ }{{\bf #1} {(#2)} {#3}}}
\nc{\jpc}[3]{{\it  J.\ Phys.\ C\ }{{\bf #1} {(#2)} {#3}}}
\nc{\jap}[3]{{\it J.\ Appl.\ Phys.\ }{{\bf #1} {(#2)} {#3}}}
\nc{\jpsj}[3]{{\it J.\ Phys.\ Soc.\ Japan\ }{{\bf #1} {(#2)} {#3}}}
\nc{\lmp}[3]{{\it Lett.\ Math.\ Phys.\ }{{\bf #1} {(#2)} {#3}}}
\nc{\mpl}[3]{{\it  Mod.\ Phys.\ Lett.\ }{{\bf #1} {(#2)} {#3}}}
\nc{\ncim}[3]{{\it  Nuov.\ Cim.\ }{{\bf #1} {(#2)} {#3}}}
\nc{\np}[3]{{\it  Nucl.\ Phys.\ }{{\bf #1} {(#2)} {#3}}}
\nc{\pr}[3]{{\it Phys.\ Rev.\ }{{\bf #1} {(#2)} {#3}}}
\nc{\pra}[3]{{\it  Phys.\ Rev.\ A\ }{{\bf #1} {(#2)} {#3}}}
\nc{\prb}[3]{{\it  Phys.\ Rev.\ B\ }{{{\bf #1} {(#2)} {#3}}}}
\nc{\prc}[3]{{\it  Phys.\ Rev.\ C\ }{{\bf #1} {(#2)} {#3}}}
\nc{\prd}[3]{{\it  Phys.\ Rev.\ D\ }{{\bf #1} {(#2)} {#3}}}
\nc{\prl}[3]{{\it Phys\ Rev.\ Lett.\ }{{\bf #1} {(#2)} {#3}}}
\nc{\pl}[3]{{\it  Phys.\ Lett.\ }{{\bf #1} {(#2)} {#3}}}
\nc{\prep}[3]{{\it Phys\. Rep.\ }{{\bf #1} {(#2)} {#3}}}
\nc{\prsl}[3]{{\it Proc.\ R.\ Soc.\ London\ }{{\bf #1} {(#2)} {#3}}}
\nc{\ptp}[3]{{\it  Prog.\ Theor.\ Phys.\ }{{\bf #1} {(#2)} {#3}}}
\nc{\ptps}[3]{{\it  Prog\ Theor.\ Phys.\ suppl.\ }{{\bf #1} {(#2)} {#3}}}
\nc{\physa}[3]{{\it  Physica\ A\ }{{\bf #1} {(#2)} {#3}}}
\nc{\physb}[3]{{\it  Physica\ B\ }{{\bf #1} {(#2)} {#3}}}
\nc{\phys}[3]{{\it Physica\ }{{\bf #1} {(#2)} {#3}}}
\nc{\rmp}[3]{{\it  Rev.\ Mod.\ Phys.\ }{{\bf #1} {(#2)} {#3}}}
\nc{\rpp}[3]{{\it Rep.\ Prog.\ Phys.\ }{{\bf #1} {(#2)} {#3}}}
\nc{\sjnp}[3]{{\it Sov.\ J.\ Nucl.\ Phys.\ }{{\bf #1} {(#2)} {#3}}}
\nc{\spjetp}[3]{{\it Sov.\ Phys.\ JETP\ }{{\bf #1} {(#2)} {#3}}}
\nc{\yf}[3]{{\it Yad.\ Fiz.\ }{{\bf #1} {(#2)} {#3}}}
\nc{\zetp}[3]{{\it Zh.\ Eksp.\ Teor.\ Fiz.\  }{{\bf #1}  {(#2)} {#3}}}
\nc{\zp}[3]{{\it Z.\ Phys.\ }{{\bf #1} {(#2)} {#3}}}
\nc{\ibid}[3]{{\sl ibid.\ }{{\bf #1} {#2} {#3}}}

\nc{\rf}[1]{(\ref{#1})}
\nc{\nn}{\nonumber \\*}
\nc{\bfB}{\bf{B}}
\nc{\bfv}{\bf{v}}
\nc{\bfx}{\bf{x}}
\nc{\bfy}{\bf{y}}
\nc{\vx}{\vec{x}}
\nc{\vy}{\vec{y}}
\nc{\oB}{\overline{B}}
\nc{\oI}{\overline{I}}
\nc{\oR}{\overline{R}}
\nc{\rar}{\rightarrow}
\nc{\ti}{\times}
\nc{\slsh}{\hskip-5pt/}
\nc{\sm}{Standard~Model~}
\nc{\MP}{M_{\rm Pl}}
\nc{\tp}{t_{\rm Pl}}

\nc{\pmin}{p_{\rm min}}
\nc{\pmax}{p_{\rm max}}
\nc{\fo}{f_0}
\nc{\foi}{f_{0,i}\,}
\nc{\fop}{f_0^P}
\nc{\fou}{f_0^U}

\nc{\eff}{{\rm eff}}
\nc{\MT}{M_{\rm T}}
\nc{\ML}{M_{\rm L}}
\nc{\kk}{\vek{k}}
\nc{\pp}{{\rm p}}
\nc{\pt}{\partial_t}
\nc{\half}{{1\over 2}}
\nc{\w}{\omega}
\nc{\uhat}{\hat{U}_\w}

\nc{\etal}{\mbox{\it et al.}}
\nc{\ie}{{\it i.e. }}
\nc{\eg}{{\it e.g. }}
\nc{\trh}{T_{\rm RH}}

\begin{document}
\title{{\hfill {{\small  TURKU-FL-P39-02
        }}\vskip 1truecm}
{\LARGE Simulations of Q-Ball Formation}
\vspace{-.2cm}}

\author{{\sc Tuomas Multam\"aki\footnote{email: tuomul@utu.fi}}\\ 
and\\
{\sc Iiro Vilja\footnote{email: vilja@utu.fi}}\\ 
\\{\sl\small Department of Physics, University of Turku, FIN-20014, FINLAND}}

\date{}
\maketitle

\abstract{The fragmentation of the Affleck-Dine condensate is studied
by utilizing 3+1 dimensional numerical simulations. The 3+1 dimensional
simulations confirm that
the fragmentation process is very similar to the results obtained
by 2+1 dimensional simulations. We find, however, that
the average size of Q-balls in 3+1 dimensions is somewhat larger that
in 2+1 dimensions. A filament type structure
in the charge density is observed during the fragmentation process.
The resulting final Q-ball distribution is strongly dependent 
on the initial conditions of the condensate and approaches a thermal
one as the energy-charge ratio of the Affleck-Dine condensate
increases.}

\newpage

\section{Introduction}

The Affleck-Dine condensate \cite{ad} offers an interesting possibility for 
baryogenesis at a high energy scale. The Minimal Supersymmetric Standard
Model (MSSM) has several flat directions in its scalar potential along
which squark, slepton and higgs fields fluctuate during inflation forming
AD-condensates \cite{dine}. The formation of an Affleck-Dine condensate is then
natural in theories with supersymmetry. 

The evolution of the AD-condensate is not as simple as one might
envision at first, but includes highly nonlinear dynamics. In 
\cite{shapo418,kari425}
it was realized that the condensate will fragment into Q-balls \cite{cole2}
which represent the true ground state instead of the condensate. This
fragmentation process has then been studied in a number
of works \cite{kasuya1}-\cite{2dsims}, both in the case of gravity and 
gauge mediated
supersymmetry breaking. Three dimensional simulations have been performed
in both cases \cite{kasuya1,kasuya2}, but only for a very limited range 
of the energy-charge
ratio of the condensate. In two dimensions, Q-ball formation has been
studied in detail in \cite{2dsims}, where it was found that 
at high initial value of the energy-charge ratio of the condensate,
the resulting Q-ball charge distribution is well represented by
a thermal distribution. On the
basis of the analytical arguments presented in \cite{2dsims}, it is
expected that the three dimensional results should be well 
represented by the two dimensional simulations. In order to verify this
assumption and to study the fragmentation process in more detail,
we have studied the fragmentation process by utilizing
full three dimensional simulations.

In this letter we present results from full 3+1 dimensional simulations
of the fragmentation of the Affleck-Dine condensate in the MSSM
with gravity mediated supersymmetry breaking. The numerical 
results are presented in Section 2 along with illustrations of
the fragmentation process. In Section 3 we study the Q-ball
distribution statistically. The letter is concluded in Section 4.

\section{Numerical results}
\subsection{Preliminaries}
The numerical work is performed by utilizing the parallel computing resources
available at the Finnish center for high-performance computing (CSC).
The simulations are run on a Cray T3E parallel machine.

In the simulations, the lattice is divided into cubes of $40^3$ and 
data transfer between the cubes is achieved by using the Message 
Passing Interface (MPI)-library. The lattice size used was typically 
$120^3$ but also larger lattice sizes up to $200^3$ were studied.
In all simulations, periodic boundary conditions were used.

As an initial condition we add uniform noise to the amplitude and the phase
of the uniformly rotating condensate field,
\be{ini}
\Phi=\Phi_0 \textrm{e}^{i\w t}+\delta\Phi.
\ee
The amplitude of the condensate is chosen to have the same value as in 
the 2D-case, $\Phi_0=10^9\GeV$. The type of noise added does not 
affect the results \cite{pkasuya}. The amplitude of the noise was
typically $|\delta\Phi|/\Phi_0\sim 10^{-10}$, but it was also varied
in order to study its effects on the results. 
Decreasing the amplitude of the noise simply delays the beginning of
the fragmentation process but the essential features of the dynamics
are unchanged.

The simulations are done with varying values of $\w$ to study the effect
of varying the initial energy-charge -ratio, $x\equiv \rho/(mq)$,
of the condensate. As $\w$ is decreased the condensate rotates 
more slowly and $x$ increases. Assuming that $V(\Phi_0)\ll\w^2|\Phi_0|^2$,
$x$ and $\w$ are related simply by $x\approx m\w^{-1}$.

The equation of motion of the AD-condensate is
\be{adeqm}
\ddot{\Phi}+3H\dot{\Phi}-{1\over a^2}\nabla^2\Phi+ 
m^2\Phi[1+K\log({|\Phi|^2\over M^2})]-g H^2\Phi+
{3\lambda^2\over\MP^2}|\Phi|^4\Phi=0, 
\ee
where $M=10^{14} \GeV$, $a$ is the scale factor of the 
universe and $H$ is the Hubble parameter, $H={\dot{a}/a}$.
The field was decomposed into real and imaginary parts,
$\Phi=\phi_1+i\phi_2$. Field and space-time were also rescaled 
according to
\be{rescalings}
\varphi={\phi\over m},\ \ h={H\over m},\ \ \tau=mt,\ \ \xi=mx.
\ee
The parameter values chosen for the simulations were
$m=10^2\GeV,\ K=-0.1$ and $\lambda={1\over 2}$ (the $g H^2$-term can be
omitted since $g H^2\ll m^2$). The universe is assumed to be matter
dominated so that $a=a_0t^{2\over 3}$ and $H={2\over 3t}$. 

The spatial and temporal lattice units were varied in order to verify that
their exact values do not affect the results.
The calculations were done in a
comoving volume so that the physical size of the lattice
increases with time. The initial time is $\tau_0=100$ and
$a=0.1\times\tau^{2\over 3}$. The simulations were run up to
$\sim 10^6$ time steps.

\subsection{Results}

We have simulated the evolution of the AD-condensate for different values of
$x$ (or $\w$) to check for similar and differing features 
in two and three dimensions. 
The increase of phase
space and somewhat larger dissipation in three dimensions are expected
to modify some features of final Q-ball systems.

We have presented the essential features of the simulations in Figs. 
\ref{fig1}-\ref{fig6}. Figs. \ref{fig1}-\ref{fig3} illustrate the evolution
condensate with $x\approx 1$, while Figs. \ref{fig4}-\ref{fig6} show the
$x\approx 10^5$-case\footnote{Color versions of the figures are available at
www.utu.fi/\tilde tuomul/qballs/}. Here we have plotted just one $40^3$-box from
the whole lattice. Note that we plot comoving volume,\ie the
physical size of the box grows with time. In the Figs. 
\ref{fig1}-\ref{fig6}, $\rho_\phi^c$ indicates the absolute value of the equal
charge density surface drawn to each figure.

In both cases, the initial part of the
condensate evolution follows along similar lines:
The fastest growing mode starts to dominate the initially
stochastic perturbation spectrum, Fig. \ref{fig1}. Linear growth 
continues until non-linear effects begin to dominate and perturbations 
begin rapid growth.
The regions of largest charge density form filament type structures which
then fragment into lumps of charge, Fig. \ref{fig2}. 

If $x\approx 1$, no further qualitative development is visible after
the filaments fragment. The energetic lumps are excited Q-balls which
relax as the universe (box) expands, leaving a distribution consisting of
Q-balls.

If $x\gg 1$, a further development, however, takes place, as is
shown in Figs. \ref{fig4}-\ref{fig6}. After the filaments have fragmented
into energetic charge lumps, a rapid growth of high density regions
with negative charge takes place (negative charge is plotted as light gray
in the figures), Figs. \ref{fig4}-\ref{fig5}. After this stage a 
large number of charged lumps which carry positive and negative charge
are present in the box. The lumps relax with time and the spatial
distribution freezes as the box grows. 

Note that in the depicted $x\approx 1$ case the amplitude of the initial
perturbation is smaller than in the $x\approx 10^5$ case, which
explains why the initial charge lumps appear more quickly in the
large $x$ case. With equal initial perturbations, the fragmentation
process takes place at the same time in both cases. 

\section{Discussion}

The qualitative features in the three dimensional case are
very similar to the previously studied two dimensional case, 
which is exactly what we expect on the basis
of analytical arguments \cite{2dsims}. To verify that 
here the Q-ball anti-Q-ball distribution in the large $x$ case
is thermal, we utilize statistical means. 

The one-particle partition function for a thermal distribution is given
by
\be{thermdist}
Z_1=\integral{V_D}{} \frac{d^D{x}~d^D{p}}{(2\pi)^D}
\integral{-\infty}{\infty} dQ~e^{-\beta E+\mu Q}
\ee
where $\mu$ is the chemical potential related to the charge of Q-balls. 
The cumulative distribution function $F(Q,\mu,\beta)$ in terms
of the probability distribution function $g(Q,\mu,\beta)$ is then
\be{CDF}
F(Q,\mu,\beta)=\integral{-\infty}{Q}dQ' g(Q')=\integral{-\infty}{Q}dQ' \e^{\mu
Q'-\beta |Q'|}(\beta|Q'|+1), 
\ee
where we have omitted the fugacity of the distribution.

Before we can fit the cumulative distribution function to the
Q-ball distribution seen in the simulations, we need to 
count and classify the lumps of charge. Our approach is two-fold:
first we search for local maxima in the charge (anti-charge) density.
Then we classify the local maxima, $\phi_{max}$, by using the criteria
\be{ballcrit}
\phi_{max}>M\exp({1-(\w/m)^2\over 2|K|}),
\ee
which follows from the fact that inside a Q-ball, $V(\phi)-\w^2\phi^2<0$.
In other words we require that the complex phase of a
charge maximum must rotate quickly enough for it to be classified
as a Q-ball.  

The number of local maxima that pass the Q-ball criterion is
only a fraction of the total number, typically the number of
points is reduced by $80-90\%$.
\begin{figure}[ht]
\leavevmode
\centering
\vspace*{52mm}
\includegraphics{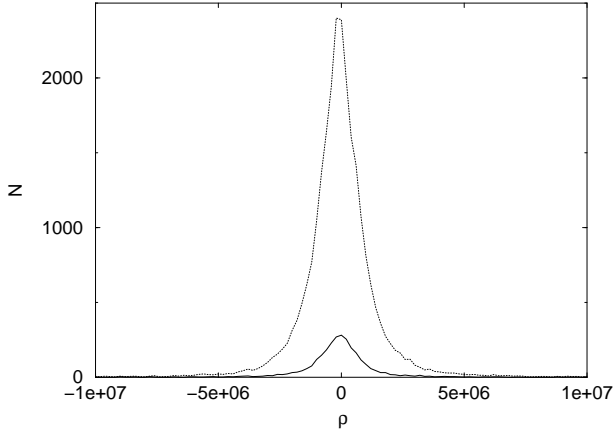}
\caption{The number of the charge maxima for $x\approx 10^5$ at 
$\tau=3\times 10^5$.}\label{histplot}
\end{figure}
This is illustrated in Fig. \ref{histplot}, where a plot 
of the local maxima is presented before (dotted line)
and after (solid line) the Q-ball criterion is applied. 
Clearly, there are both small and large charge lumps do not satisfy 
the criterion (\ref{ballcrit}). The existence of these lumps that
are not Q-balls is expected since the remains of the perturbed
initial condensate are still present in the system. Also
some of the lumps may be excited Q-balls which do not yet
satisfy the condition (\ref{ballcrit}).

\begin{figure}[ht]
\leavevmode
\centering
\vspace*{52mm}
\includegraphics{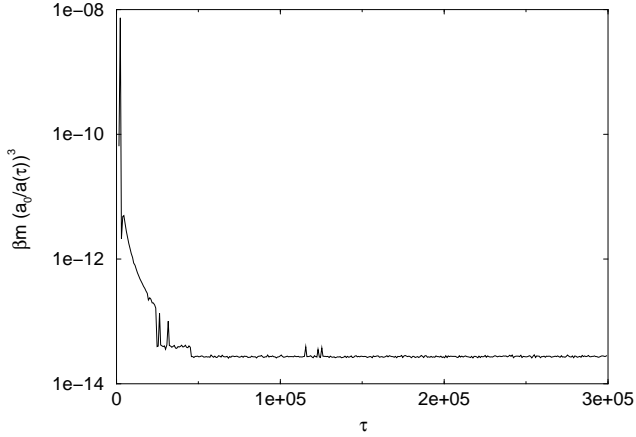}
\caption{$m\beta$ as function of $\tau$, for $x\approx 10^5$.}\label{statfig}
\end{figure}

\begin{figure}
\leavevmode
\centering
\vspace*{50mm}
\includegraphics{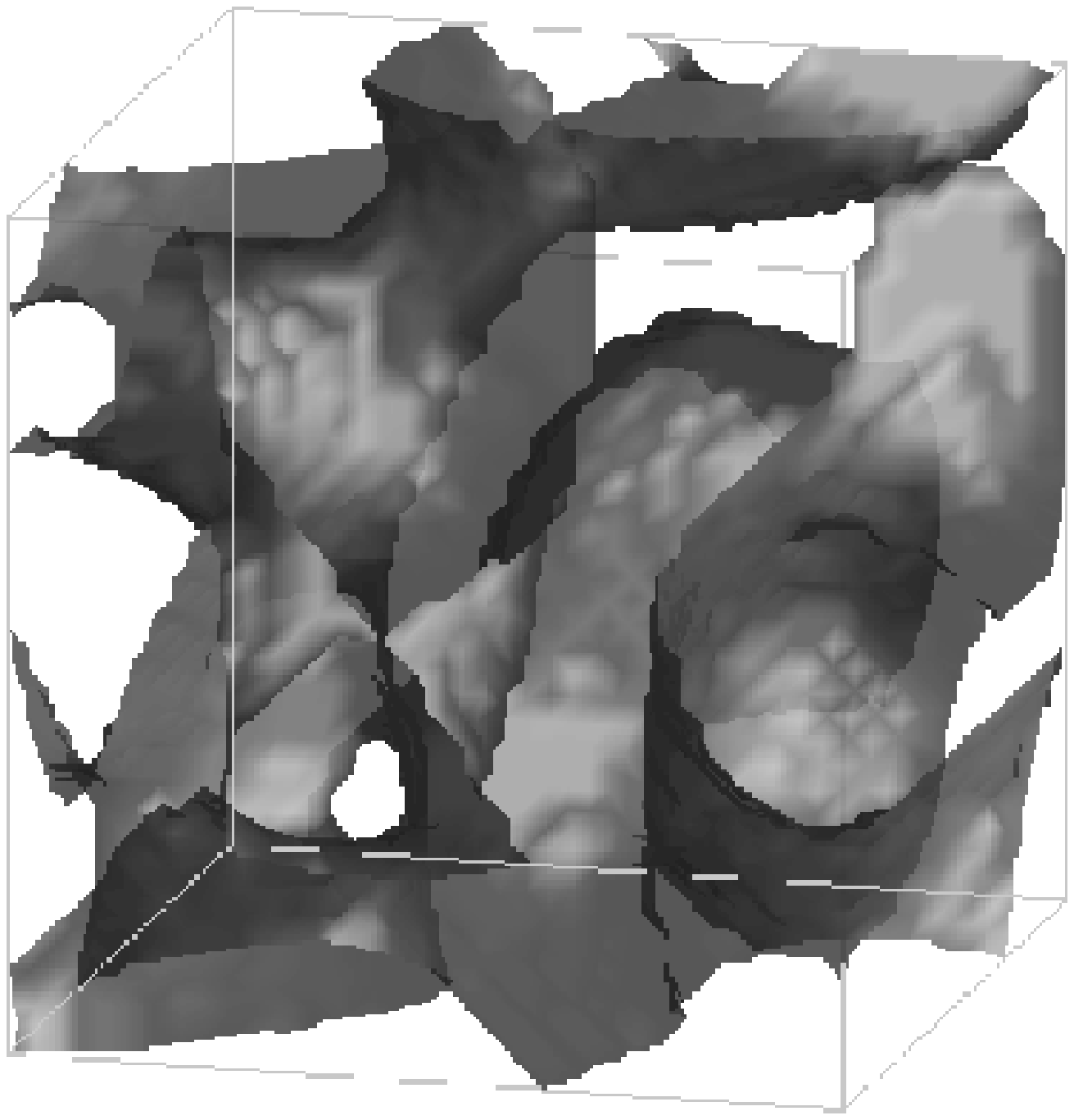}
\includegraphics{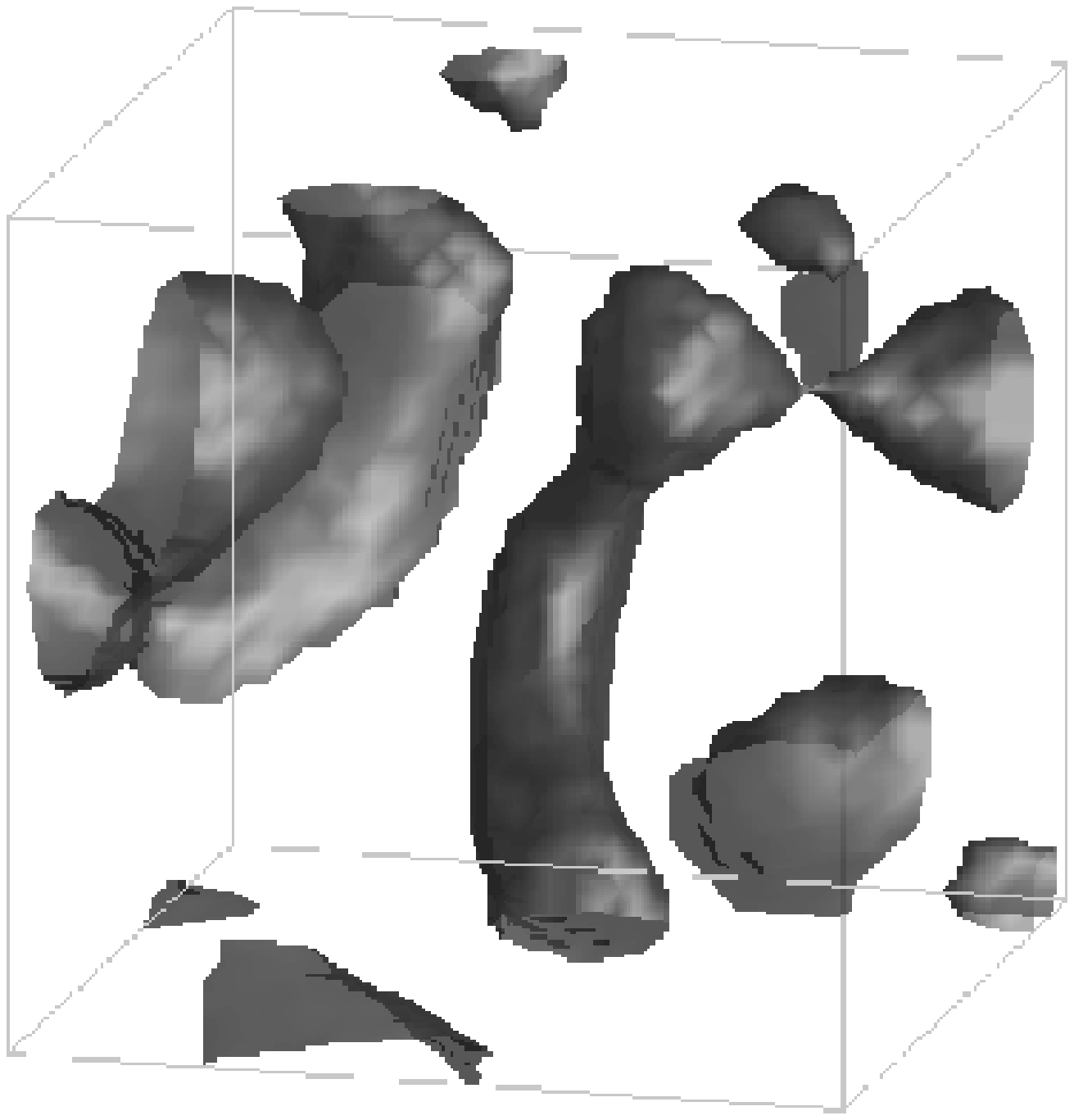}
\caption{$\tau=875\, (\rho^c_{\varphi}=10^{12}),\ 1000\, 
(\rho^c_{\varphi}=8\times10^{11})$, for $x\approx 1$.}\label{fig1}
\end{figure}

\begin{figure}
\leavevmode
\centering
\vspace*{50mm}
\includegraphics{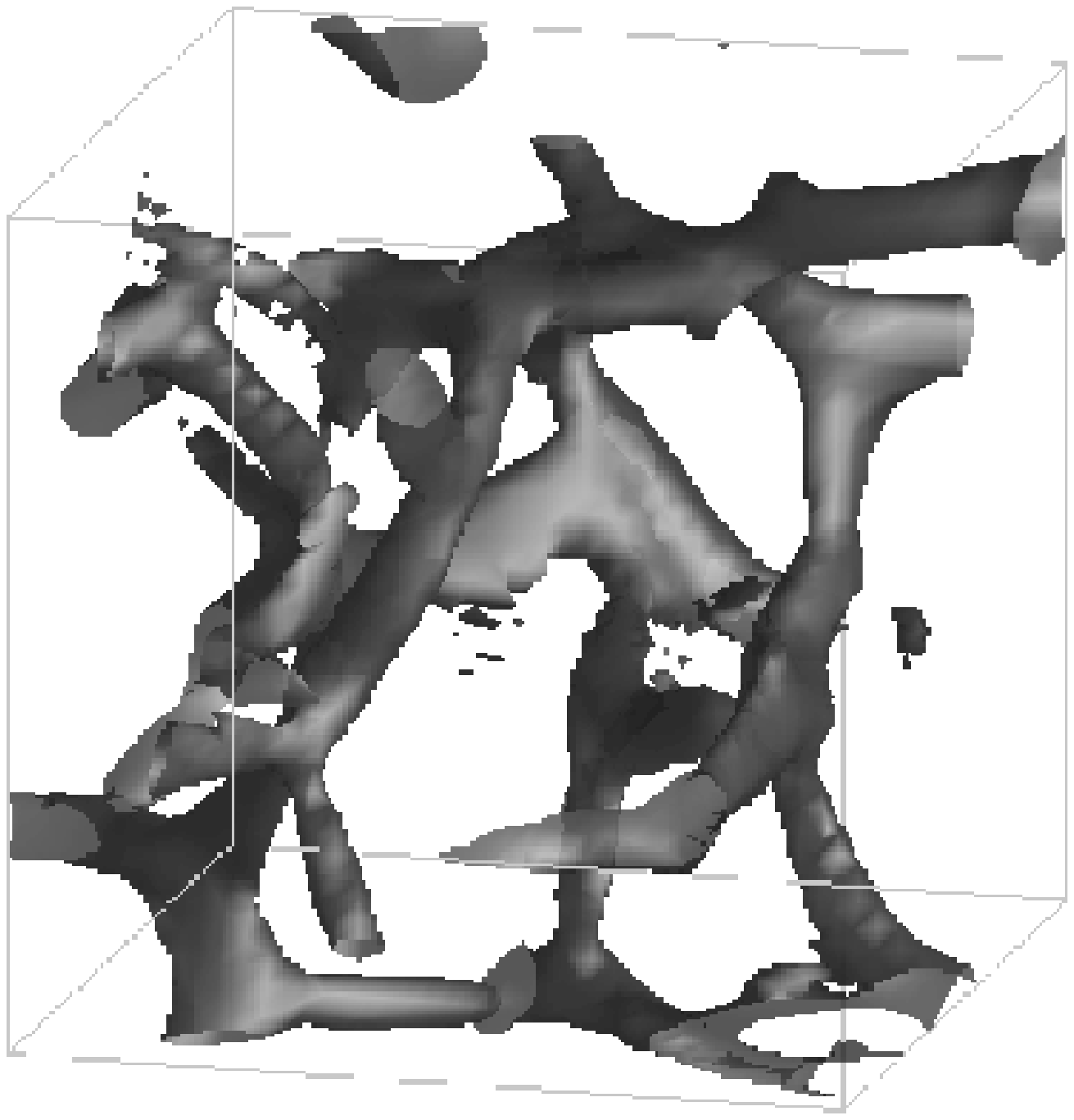}
\includegraphics{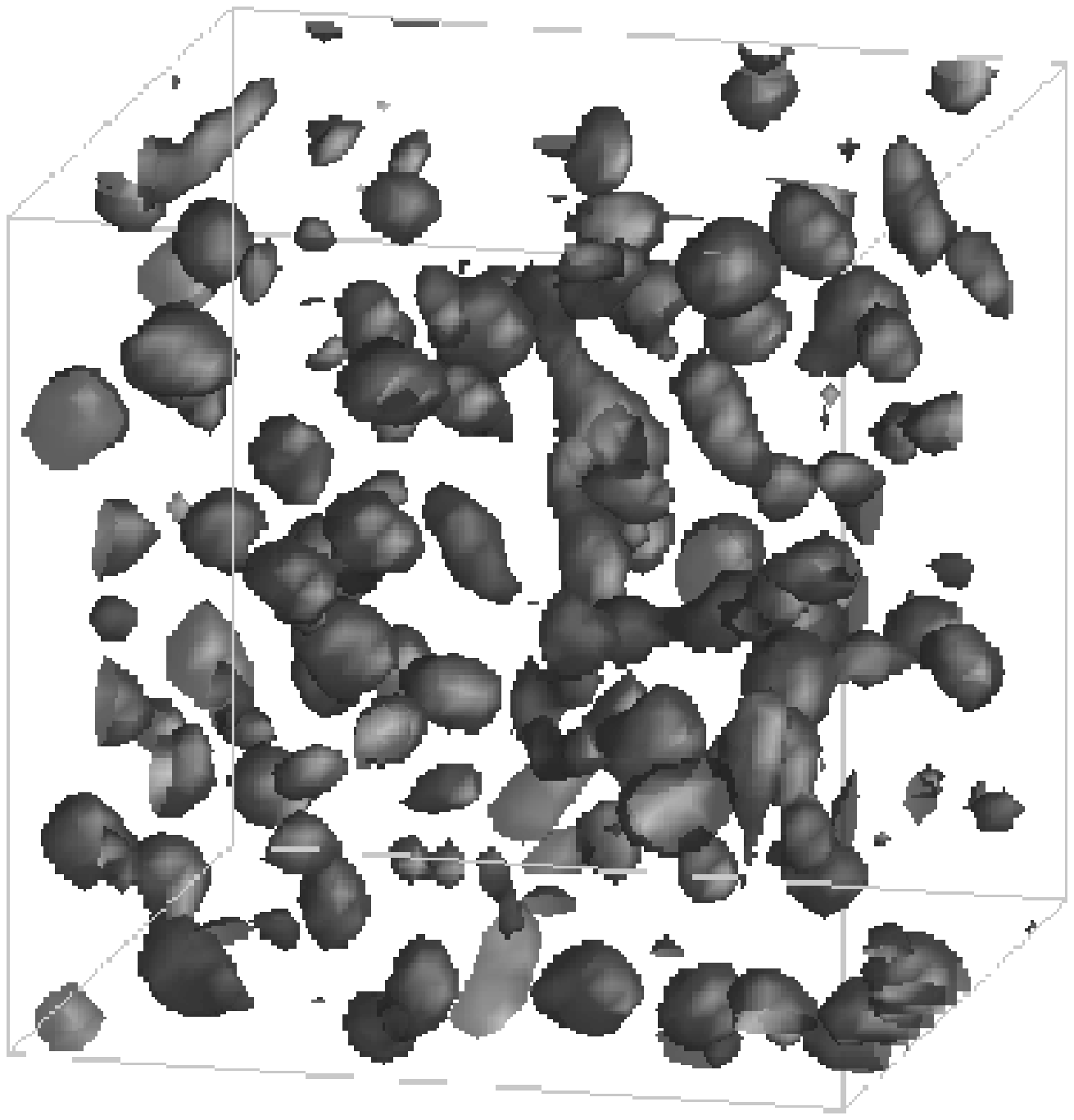}
\caption{$\tau=1875,\ 2000\, (\rho^c_{\varphi}=10^{11})$, for
$x\approx 1$.}\label{fig2}
\end{figure}

\begin{figure}
\leavevmode
\centering
\vspace*{50mm}
\includegraphics{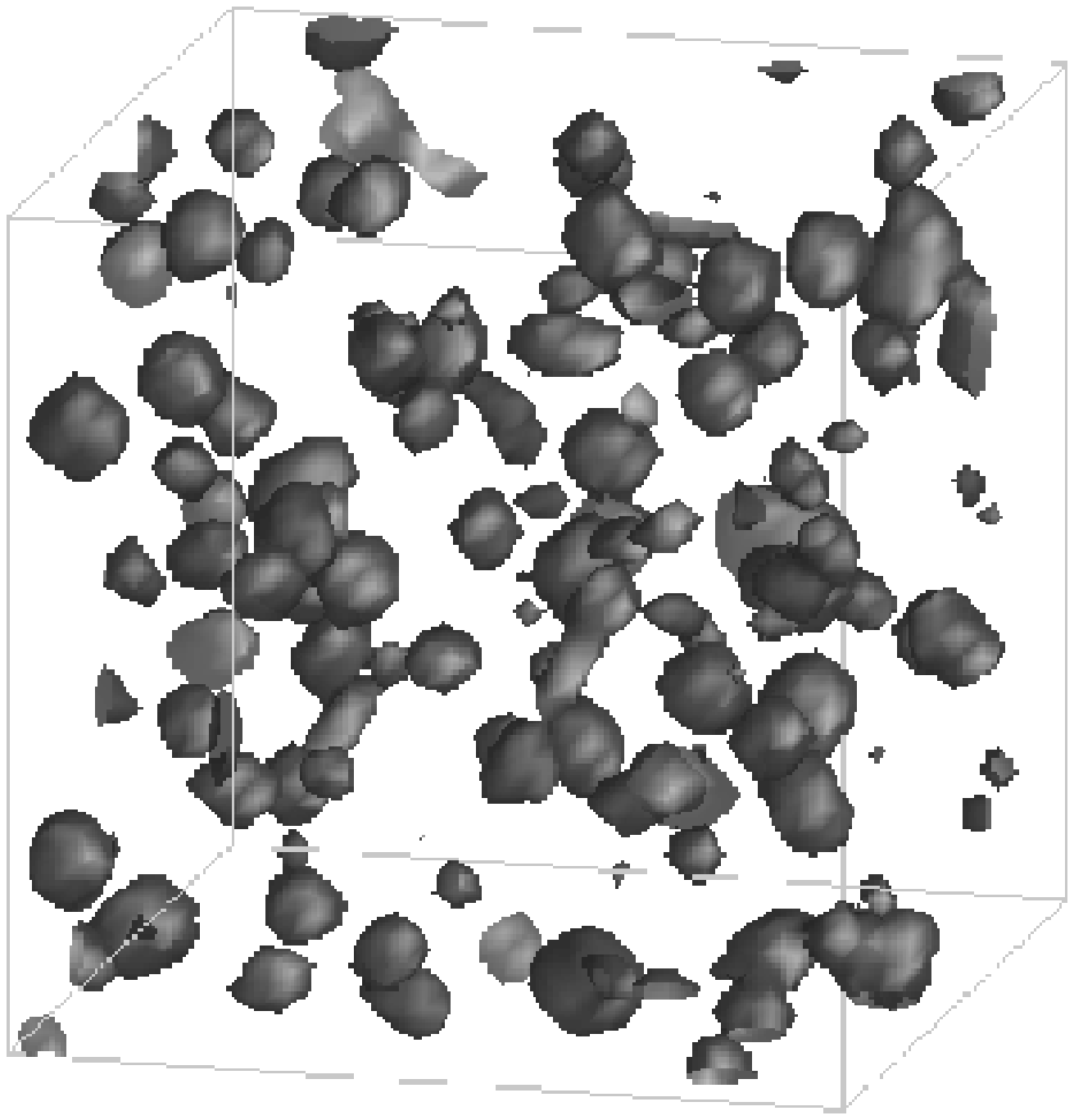}
\includegraphics{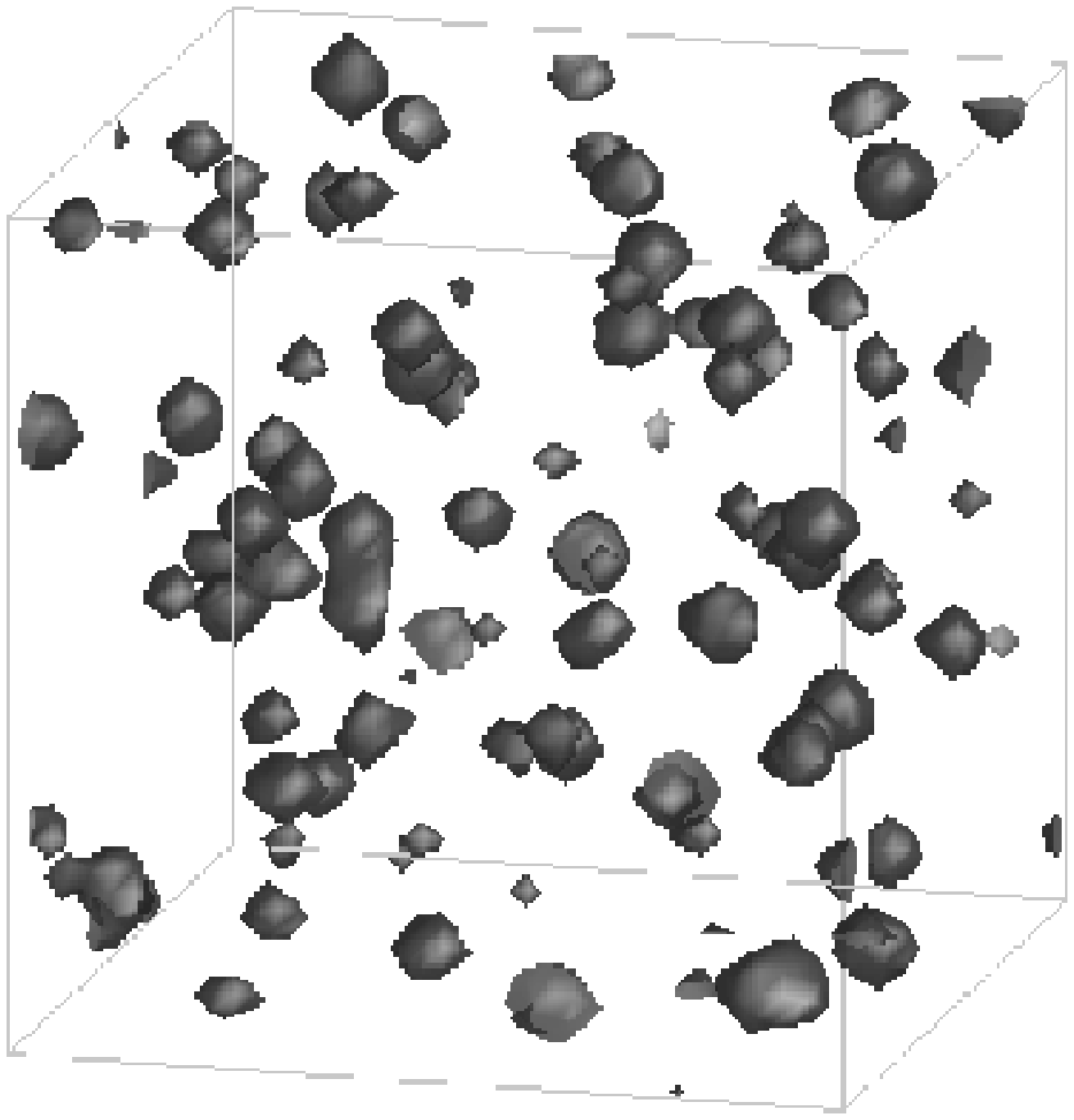}
\caption{$\tau=2250,\ 3000\,(\rho^c_{\varphi}=10^{11})$, for
$x\approx 1$.}\label{fig3}
\end{figure}

\begin{figure}
\leavevmode
\centering
\vspace*{50mm}
\includegraphics{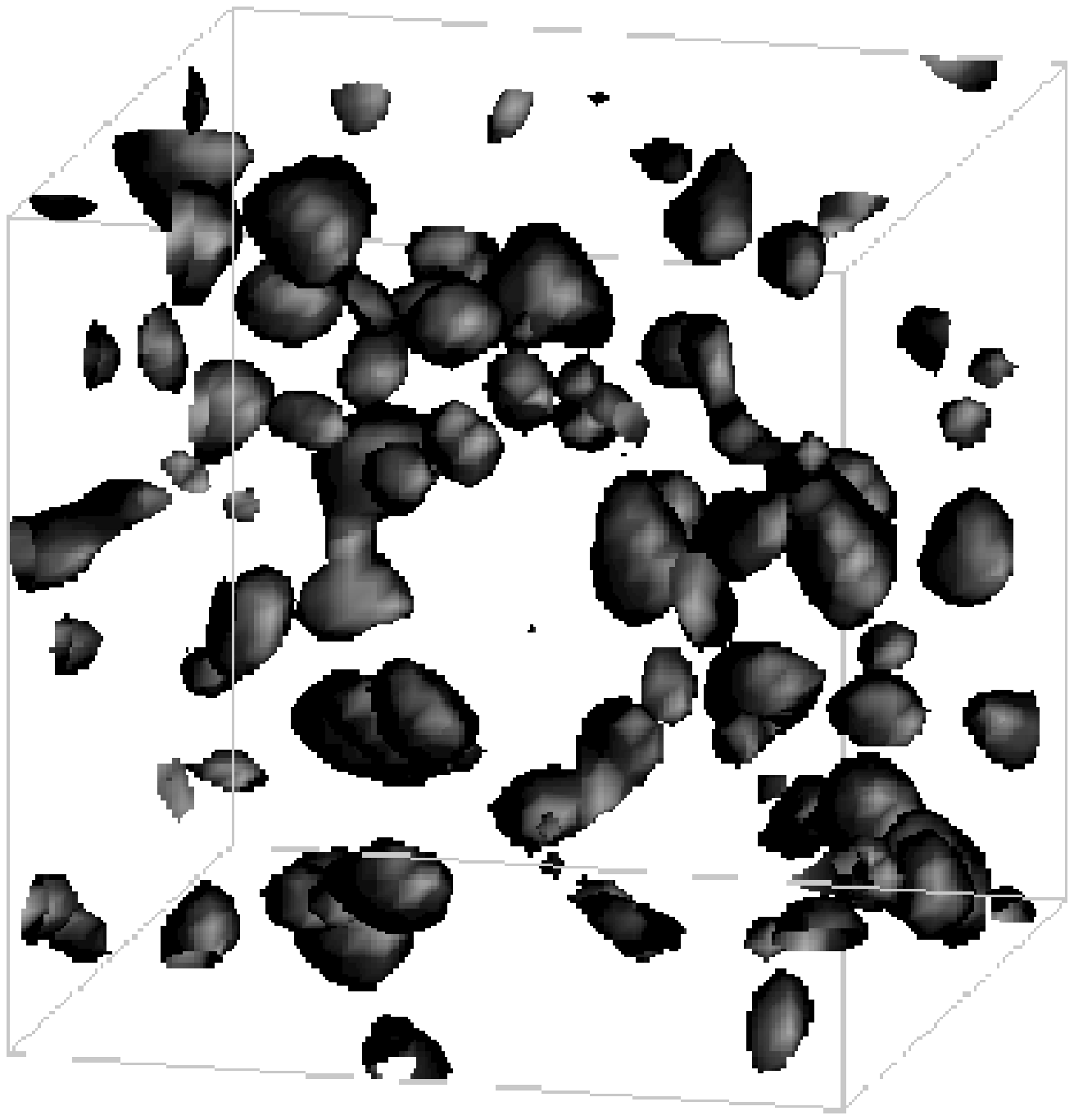}
\includegraphics{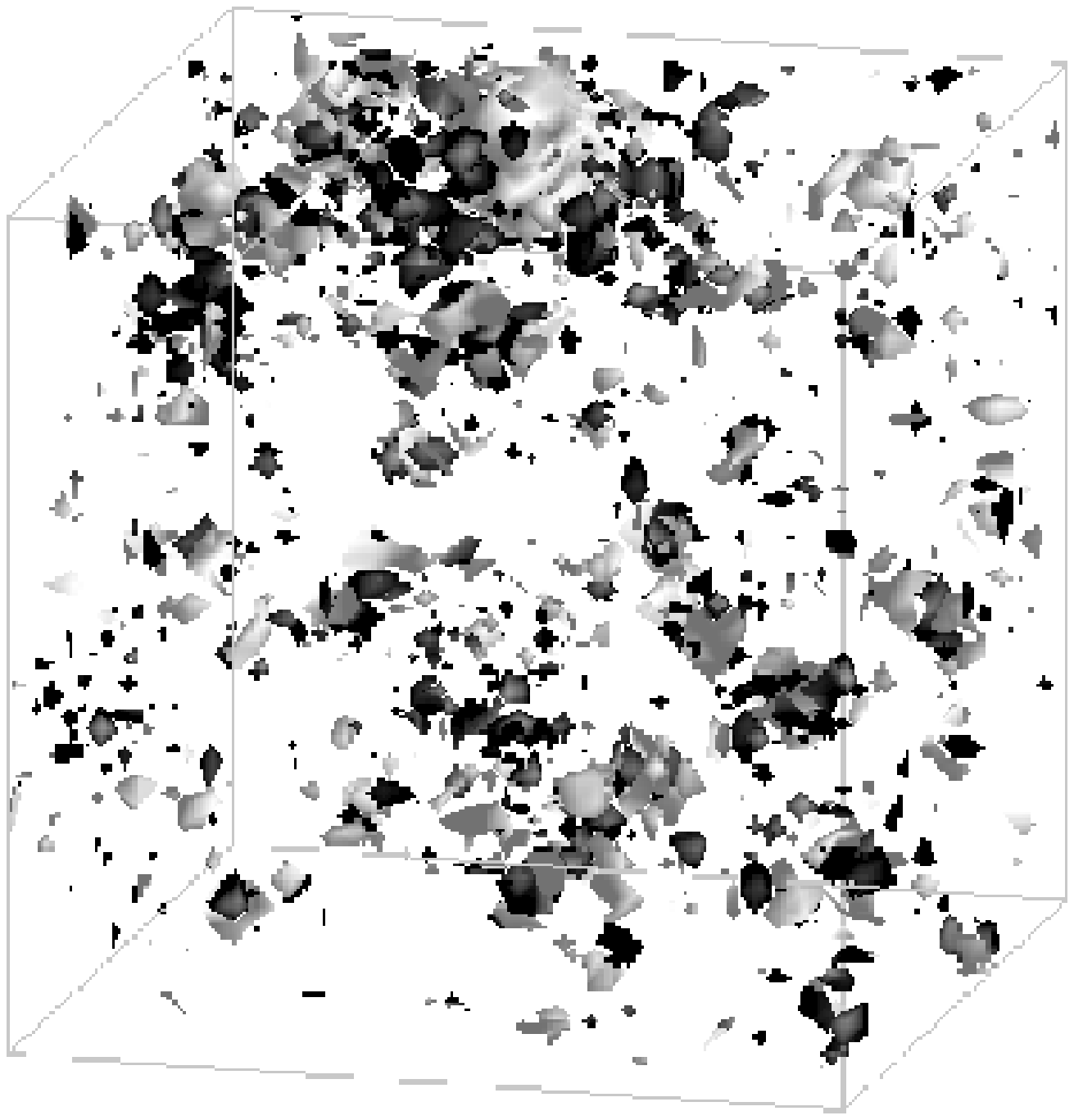}
\caption{$\tau=1500\, (\rho^c_{\varphi}=2\times10^{6}),\ 3000\, 
(\rho^c_{\varphi}=3\times10^{7})$, for $x\approx 10^5$.}\label{fig4}
\end{figure}

\begin{figure}
\leavevmode
\centering
\vspace*{50mm}
\includegraphics{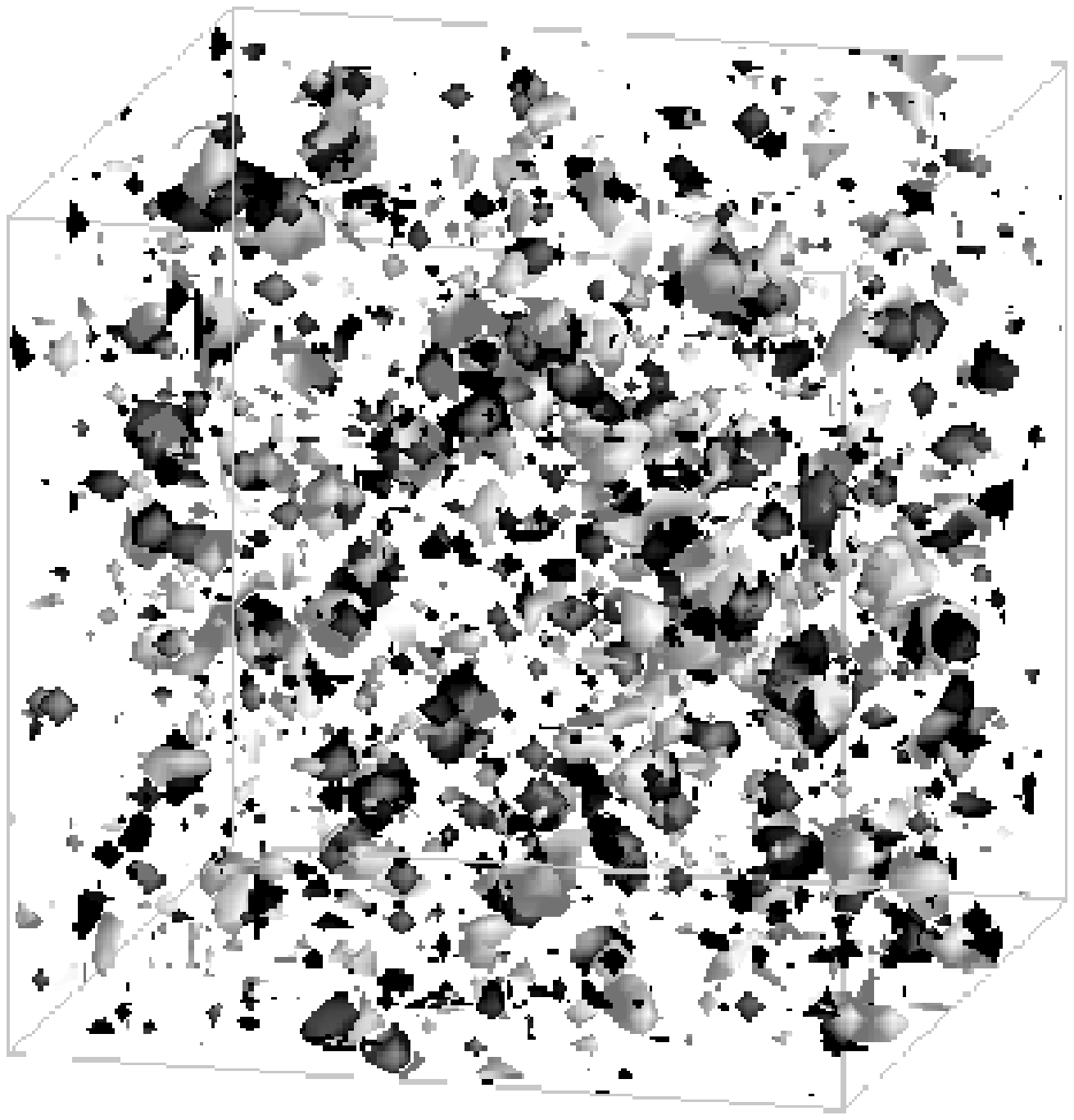}
\includegraphics{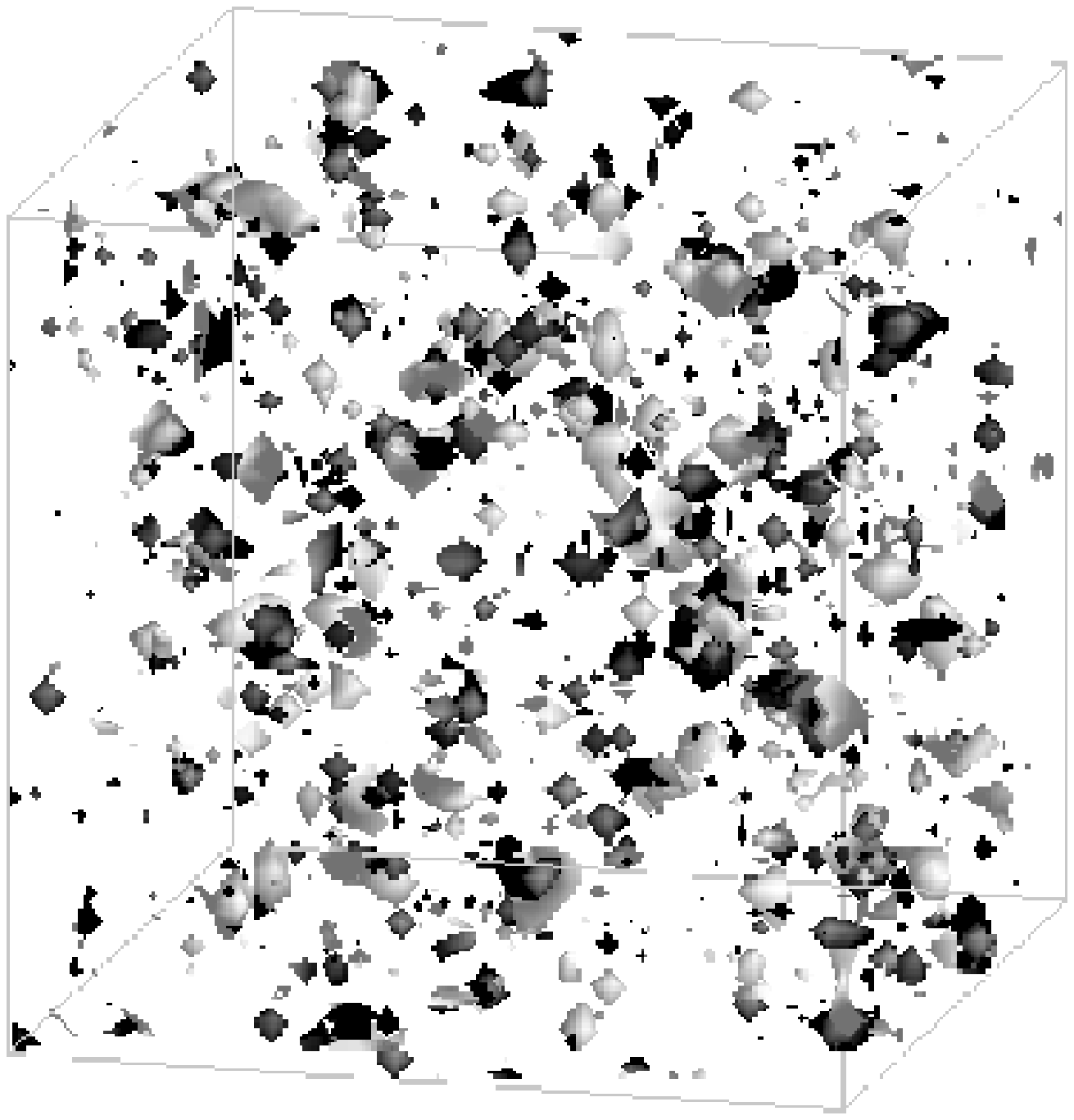}
\caption{$\tau=4500,\ 5250\, 
(\rho^c_{\varphi}=8\times10^{8})$, for $x\approx 10^5$.}\label{fig5}
\end{figure}

\begin{figure}
\leavevmode
\centering
\vspace*{50mm}
\includegraphics{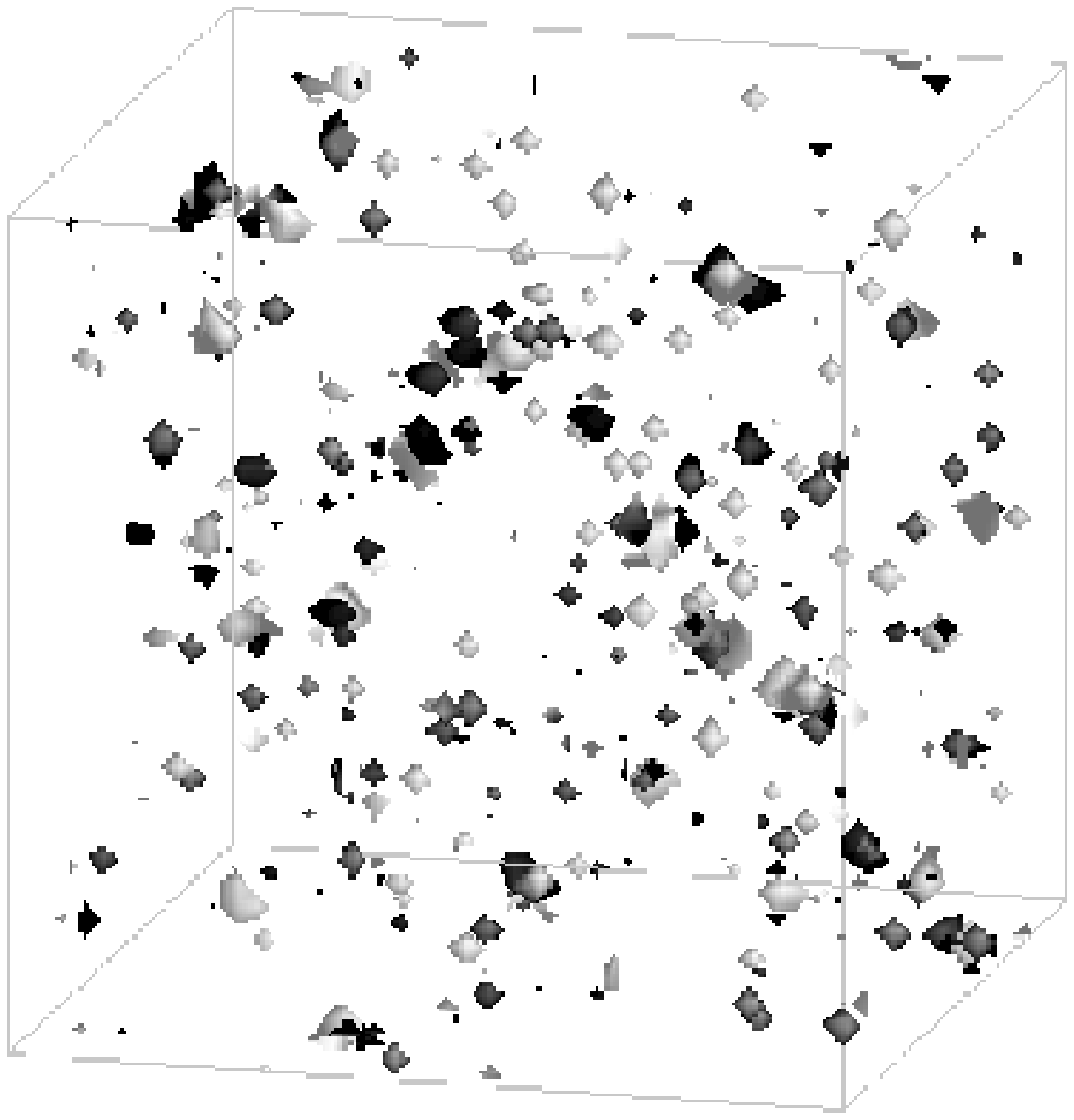}
\includegraphics{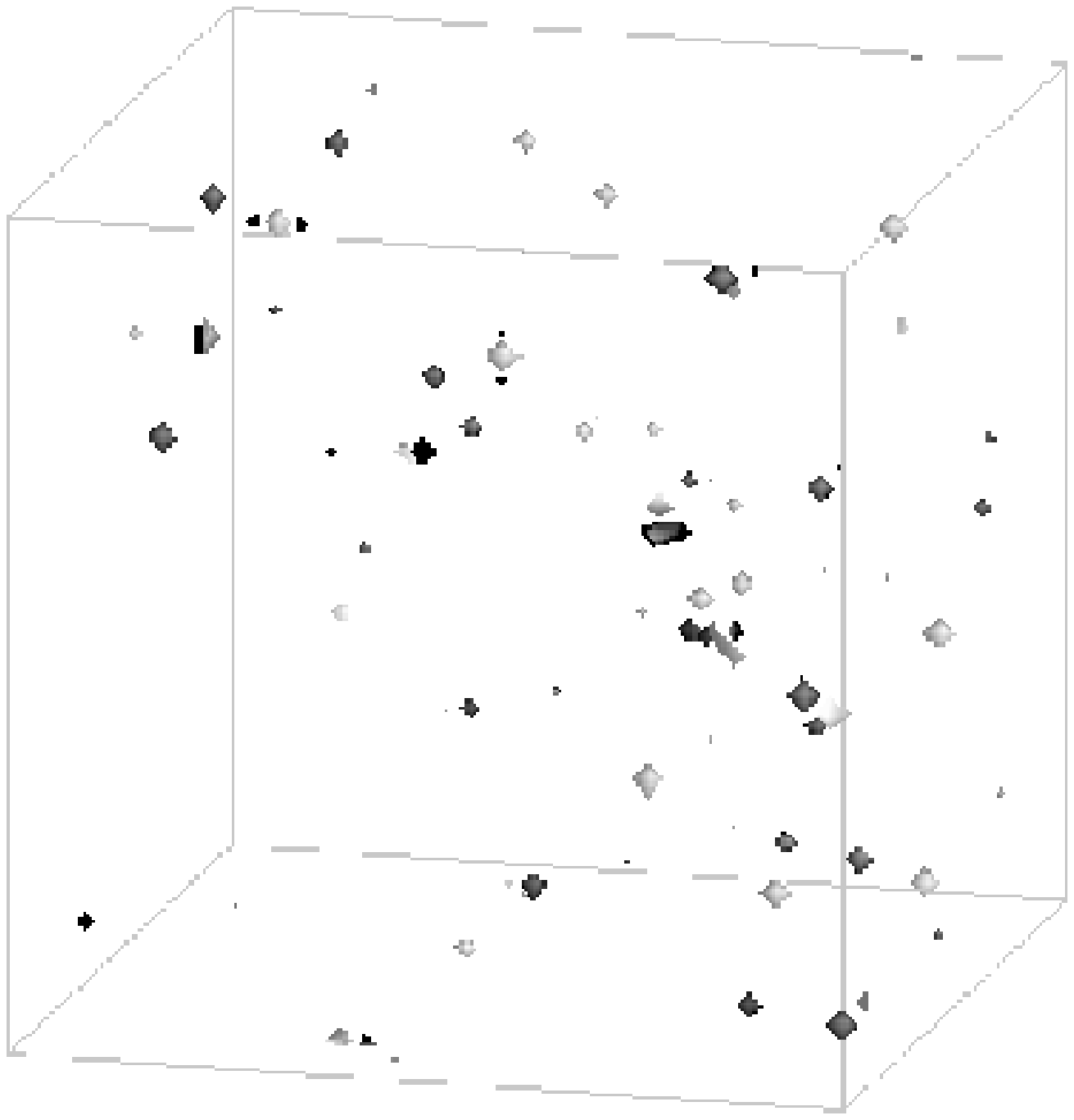}
\caption{$\tau=7500,\ 1.5\times 10^{4}\, (\rho^c_{\varphi}=8\times 10^{8})$, 
for $x\approx 10^5$.}
\label{fig6}
\end{figure}

After selecting which charge lumps to consider, we then fit the
cumulative distribution function to the Q-ball charge distribution.
The values of $m \beta$ at different $\tau$ are shown in Fig. \ref{statfig}
(corrected for the expansion of the box).
Just like in the two dimensional case, we see that the Q-ball
anti-Q-ball -system reaches a thermal equilibrium soon after 
negative charge enters the system. The chemical potential, $\mu$,
is also again much smaller than $m\beta$. Due to geometrical 
considerations, the number density of Q-balls
is slightly larger in three dimensions. Because the radii of
Q-balls are essentially equal in both dimensions \cite{MV},
the higher the dimension is, the more Q-balls the system can support.

If the criterion (\ref{ballcrit}) is not applied and (\ref{CDF})
is fitted to the whole distribution, we find that the value of
$m\beta$ is practically unchanged. On this basis, we expect that
the value of $m\beta$ obtained from the whole distribution in 
two dimensions \cite{2dsims}, is a good approximation to the 
distribution with the Q-ball criterion applied.

The value of $m\beta$ in the three dimensional simulations performed here is 
smaller than in the two dimensional case \ie 
the average energy of a Q-ball is larger. 
One can understand this by considering the total energies
in the system in the two cases (omitting $\mu$) \cite{2dsims},
$E_D=(D+1)N_D/\beta_D$, where $D$ is the dimension and $N$
is the number of Q-balls. From this we get
$\beta_{3}=(4 E_2 N_3)/(3 E_3 N_2)\beta_{2}$.
The energies in the two cases are related by $(E_2/E_3)=(m/\GeV)^{-1} V_2/V_3$,
where $V_D$ is the volume in $D$ dimensions in lattice units. 
The number densities
are found to be ${N_3/V_3}\approx 2 {N_2/V_2}$, hence
$\beta_3\approx 8/3 (\GeV/m)\beta_2$, which is in good agreement
with the numerical results.

\section{Conclusions}

We note that the
results fully support the expectations that the two dimensional
simulations capture the essential features of the fragmentation
process, both at qualitative and quantitative level.
The final Q-ball distribution is strongly dependent on the initial
charge-energy ratio $x$: with $x\approx 1$, only positive charge
appears due to the fact that Q-balls in this scenario have a linear
energy dependence on charge, $E\approx m Q$ \cite{kari538}, so that all energy
and charge can be accounted for by creating Q-balls.
As $x$ grows, the excess energy left after storing the charge
in Q-balls grows and, as it must be accounted for, leads to
the production of negative charge. In the $x\approx 10^5$ case, we
have again seen that $|Q_+|+|Q_-|>>Q_++Q_-$ \ie that the
absolute amount of charge created is much larger than the
net charge in the box. This fact also presents itself in
the form of a small chemical potential $\mu$.

The charge distribution in the large $x$ case is, like in
two dimensions, well represented by a thermal distribution.
After the initial formation process, the distribution thermalizes
slightly more slowly than in two dimensions, which is
natural due to the increased phase space. The value of $\beta$
is also compatible with the analytical and numerical expectations.

An interesting feature seen in the simulations is the formation of
a filament network in early stages of the condensate evolution. 
This was also seen in two dimensions but the observation of this
effect here confirms that this is not an artifact of two dimensions. 
Numerical analysis in two dimensions also showed that the
final distribution of large Q-balls was not spatially random. 
Small Q-balls have a large kinetic energy component, which effectively
will randomize the distribution, but the spatial distribution of 
large Q-balls may well have some structure also in three dimensions.
If Q-balls survive thermal erosion
\cite{kari538,tuomas} and decay into baryons after the electroweak
phase transition, some remains of this structure 
may still be present during nucleosynthesis. Note, however, that
any remaining spatial correlation is very small: 
assuming a high reheat temperature,
the universe can expand between reheating and nucleosynthesis by
a factor of $\sim 10^8$. Further assuming a large correlation
length of $\sim 10\ \GeV^{-1}$ (the size of the box when filaments
appear), we see that during nucleosynthesis, spatial correlations 
are very small unless the universe expands significantly between 
fragmentation and reheating.

To conclude, we have verified that the two dimensional simulations
capture all of the essential features of the dynamics of the
Affleck-Dine condensate. Depending on the initial conditions, 
the AD-condensate fragments into a large number of Q- or anti-Q-balls.
A period during which the charge density forms filament like structures
is observed, which is likely to leave a spatial correlation between large
Q-balls.


\section*{Acknowledgments}
We thank S. Kasuya for discussions.
This work has been partly supported by the Magnus Ehrnrooth Foundation.
The Finnish center for high-performance computing and
networking (CSC) is gratefully acknowledged for providing the 
computational resources.

\newpage

\newpage

\end{document}